# Strong authentication on smart wireless devices

P. Caballero-Gil
Department of Statistics, OR and Computing. University of La Laguna. Spain
pcaballe@ull.es

V. Mora-Afonso
Department of Statistics, OR and Computing. University of La Laguna. Spain
alu3966@etsii.ull.es

J. Molina-Gil
Department of Statistics, OR and Computing. University of La Laguna. Spain
jmmolina@ull.es

## ABSTRACT
The rapid deployment of wireless technologies has given rise to the current situation where mobile phones and other wireless devices have become essential elements in all types of activities, including in the home. In particular, smartphones and laptops are used for wirelessly sharing photos and documents, playing games, browsing websites, and viewing multimedia, for example. This work describes a proposal for both desktop and mobile applications that use Identity-Based Cryptography (IBC) to protect communications between smart wireless devices in the home. It combines the use of IBC for Wi-Fi and Bluetooth communication, with the promising Near Field Communication (NFC) technology for secure authentication. The proposed scheme involves NFC pairing to establish as public key a piece of information linked to the device, such as a phone number or an IP address. In this way, such information can be then used in an IBC scheme for peer-to-peer communication. This is a work in progress, but preliminary implementations of prototypes on several mobile platforms have already produced promising results.

## Categories and Subject Descriptors
C.2.0 [**Computer-Communication Networks**]: General – Security and protection.

D.4.6 [**Operating Systems**]: Security and Protection Access controls – Authentication.

K.6.5 [**Management Of Computing And Information Systems**]: Security and Protection – Authentication.

## 1. INTRODUCTION
The use of wireless devices such as smartphones and laptops has been increasing rapidly worldwide for the last years in such a way that they have overtaken desktop computers. Moreover, the figures say that this tendency is to be accelerated in the next future [1]. Wireless devices are usually interconnected and can be found in all types of environments, such as homes, offices, public spaces, transportation, etc. They are being used more and more both to acquire and to apply knowledge about the environment and its users in order to improve user experience in that environment. However, in order to implement these applications safely, strong authentication allowing interaction between different devices using several technologies is necessary.

The most widely used smart devices in the home are smartphones and laptops. Both are wireless and mobile devices that are always connected, can operate to some extent autonomously, and are capable of voice and video communication, Internet browsing, and geolocation. In particular, smartphones, which are nowadays used for everything: from checking email to social networking, paying tickets, communication between peers, data sharing, etc., have lower computing capabilities than a standard computer, as well as power and battery limitations. Thus, every operation or computation in a smartphone must be implemented taking into account these constraints, maximizing efficiency and memory usage in order to avoid possible problems. Despite these difficulties, still secure communication mechanisms and protocols need to be provided in different layers and applications in order to offer security to end users in a transparent way so that they can find functional mobile applications both user-friendly and robust.

The so-called Identity-Based Cryptography (IBC) may be seen as an evolution of Public-Key Cryptography because IBC gets rid of most of the problems of traditional Public-Key Infrastructures (PKIs), as they are usually related to certificates, and certificates are no more necessary in IBC. Furthermore, it provides the same security level, but using shorter encryption keys and more efficient algorithms. In the past decade, IBC has been subject of intense study, and many different Identity-Based Encryption (IBE), Identity-Based Signature (IBS), key agreement and signcryption schemes have been proposed [3,5,9]. The characteristics of these schemes perfectly fit with the demand and requirements of not only smartphones but also other wireless devices in terms of simplicity, security, efficiency and scalability.

Near Field Communication (NFC) is a short-range high frequency wireless communication technology [13] that enables simple and safe interactions between electronic devices at a few centimeters, allowing consumers to perform contactless transactions, access digital content, and connect electronic devices with a single touch.

This work aims to propose a mobile application based on mechanisms and protocols that use Identity-Based Cryptography to provide security. In particular, it proposes a new NFC-based application for Bluetooth/Wi-Fi pairing in order to allow peer-to-peer communication whose security is based on IBC.

This paper is organized as follows. In Section 2 a brief description of related work and an introduction to some necessary cryptographic primitives are included. Then, Section 3 presents the proposed scheme and discusses its design and implementation. Finally, Section 4 provides the conclusion and future work.

## 2. BACKGROUND

### 2.1 Related Work
Not many works have proposed the use of IBC in mobile applications till now. This section briefly describes some of them.

The paper [6] proposes a special application for mobile devices that can be used to share information in groups thanks to a combination of two types of IBE, and describes its implementation for the Android operating system.

Another interesting proposal to exploit any available connection when there is extreme wireless coverage failure, also developed on the Android mobile platform, is presented in [11], where the use of IBC is suggested.

On the one hand, social networks are one of the most important application fields where IBC has been used to deal with the lack of secure mechanisms for communications among users of different social providers because if a user's account gets compromised all their contacts get indirectly compromised and could become victims of different attacks such as phishing. There are a few applications [7,10,15] that prevent users from being scammed, provide access control, etc.; but usually current security schemes of social networks assume that third parties are honest-but-curious and so they are not as scalable as current social networks demand. In [22] a security-transparent smartphone-based solution to this problem is presented. Such an approach uses IBC as mechanism to protect data, which can be applied with any third-party social network provider. An Android implementation of such a scheme is presented as a mobile application using a mobile phone text and voice messaging communication service [18], and the Pairing-Based Cryptography library [19]. Social networks are also the goal of the paper [14], which proposes a system based on IBC that enables users to communicate securely on their mobile phones using their identities and address books. This system is integrated into a social app platform that allows users to share text and photos and play games, without any trusted third party acting as an intermediary in all communications.

On the other hand, regarding access control, a framework for mobile role-based scheme using an IBE scheme with non-interactive Zero-Knowledge Proof authentication is proposed in the paper [12]. Also IBC and access control combining authentication with role-based access control is proposed in [20] to provide a cryptographic solution for large organizations.

Finally, there are several proposals related to encrypting phone calls. An Android-based product that provides encrypted phone calls using IBE is described in [21]. Such a proposal is not peer-to-peer and requires that the cell phone provider act as a trusted central server in order to manage the security of the communications. In the paper [17], an identity-based key agreement system for 2G and 3G mobile telephony is presented. There, the use of telephone numbers as public keys allows the system to increase the security of key management in the existing GSM and UMTS infrastructure.

The goal of this work is different from the ones of the aforementioned papers because here an IBC scheme is applied for peer-to-peer communication between smartphones and laptops, using as public key a piece of information linked to the device, which is established through NFC pairing.

### 2.2 Preliminaries of IBC
The notations used within this paper are summarized in Table 1.

**Table 1. Notation**

| Symbols | Meanings |
|---|---|
| $p$ | a 512-bit to 7680-bit prime number |
| $q$ | a 160-bit to 512-bit prime number |
| $F_p$ | finite field of prime characteristic p |
| $Z_p$ | additive group of integers modulo p |
| $F_p^2$ | extension field of degree 2 of the field $F_p$ |
| $E/F_p$ | elliptic curve $y^2=x^3+ax+b$ for specified a, b in $F_p$ |
| $E(F_p)$ | additive group of points of affine coordinates (x,y) with x,y in $F_p$, satisfying the curve equation $E/F_p$, including the point at infinity 0 |
| $E(F_p^2)$ | group of points of affine coordinates in $F_p^2$, satisfying the curve equation $E/F_p$, including the point at infinity 0 |
| $G_1, G_2$ | cyclic groups |
| $e$ | cryptographic bilinear map $G_1 \times G_1 \rightarrow G_2$ |
| $F^*$ | multiplicative group of the non-zero elements in the field F |
| $P$ | generator of $G_1$ |
| $P+Q$ | sum point resulting from adding two points P and Q on a curve E |
| $nP$ | multiplication point resulting from adding a point P on a curve to itself a total of n times |

Identity-Based Cryptography is a type of public-key cryptography where a public piece of information linked to a node is used as its public key. Such information may be an e-mail address, domain name, physical IP address, phone number, or a combination of any of them. Shamir described in [16] the first implementation of IBC. In particular, his proposal is an identity-based signature that allows verifying digital signatures by using only public information such as the user's identifier. Shamir also proposed identity-based encryption, which is particularly attractive because there is no need to acquire an identity's public key prior to encryption. However, he was unable to come up with a specific solution, and the definition of an IBE remained an open problem for many years. In particular, IBE may be defined as a public-key encryption scheme where any piece of information linked to an identity can act as valid public key. In IBE, a Trusted Third Party (TTP) server, the so-called Private Key Generator (PKG), first stores a secret master key used to generate a set of public parameters and the corresponding users' private keys. After a user's registration, it receives the public parameters and its private key. Then, a secure communication channel can be established without involving the PKG.

Many IBC schemes are built upon bilinear maps. A cryptographic bilinear map may de defined as a map e that satisfies the following properties:

- Bilinear: $e(aP,bQ) = e(P,Q)^{ab}$ for all $P,Q \in G_1$ and all $a,b \in Z^*_q$.
- Non-degenerate: $e(P,P)$ is a generator of $G_2$. In other words, $e(P,P) \neq 1$.
- Computable: Given $P,Q \in G_1$ there is an efficient algorithm to compute $e(P,Q)$.

Two of the best-known bilinear maps are the Weil and Tate pairings. Both are defined on the points of an elliptic curve. An algorithm for evaluating them, which is linear in the size of the input, is the so-called Miller's algorithm. Furthermore, the time that this algorithm requires to compute the pairings can be reduced if it uses a Solinas prime, which is a prime number of the form $2^a \pm 2^b \pm 1$, where $0 < b < a$.

Currently, most IBC schemes are based on assumptions of hard problems in elliptic curves. The most frequently used ones are:

- Computational Diffie-Hellman: No efficient algorithm exists to compute $abP$ from $P,aP,bP \in G_1$ for some $a,b \in Z^*_q$.
- Weak Diffie-Hellman: No efficient algorithm exists to compute $sQ$ from $P,Q,sP \in G_1$ and $s \in Z^*_q$.
- Bilinear Diffie-Hellman (BDH): No efficient algorithm exists to compute $e(P,P)^{abc}$ from $P,aP,bP,cP \in G_1$ where $a,b,c \in Z^*_q$.
- Decisional Bilinear Diffie-Hellman: No efficient algorithm exists to decide if $r=e(P,P)^{abc}$ given $P,aP,bP,cP \in G_1$, $r \in G_2$ and $a,b,c \in Z^*_q$.

Boneh and Franklin proposed in [5] the first provable secure IBE scheme. Its security is based on the hardness of the BDH problem [4]. They build the IBE system from a symmetric bilinear map e between two cyclic groups $G_1, G_2$ of order q, where $G_1$ is $E(F_p)$ and $G_2$ is a subgroup of $F^*_{p^2}$.

IBE schemes are usually divided into four main stages:

1. Setup stage. This stage is executed by the PKG just once in order to create the whole IBE environment. The master key is kept secret and used to obtain users' private keys, while the remaining system parameters are made public.
2. Extract stage. The PKG performs this stage when a user requests a private key.
3. Encrypt stage. This stage is run when any user wants to encrypt a message M to send the encrypted message C to a user whose public IDentity is ID.
4. Decrypt stage. Any user that receives an encrypted message C runs this stage to decrypt it using its private key and recovering the original message M.

Note that the verification of the authenticity of the requestor and the secure transport of the private key are problems out of the scope of usual IBE protocols. However, the authenticity of the public keys is guaranteed implicitly as long as the transport of the private keys to the corresponding user is kept secure.

## 3. NFC AUTHENTICATION FOR IBE

Experience with Bluetooth so far has shown some of its weaknesses, which should be solved for improving its use. One of its main drawbacks is the necessary pairing process [2], because pairing two unknown devices requires over ten seconds. Furthermore, such a procedure should be transparent, letting the users focus on what they want to do and not on painful side operations. In conclusion, Bluetooth pairing lacks user-friendliness. In this regard, NFC might be used for Bluetooth secure pairing [13]. In this section we describe a new scheme using NFC-based Bluetooth pairing for providing a communication channel between smartphones, which is more secure than others described in previous works based on E0 or E1 stream ciphers [23]. In particular, the scheme here described applies IBE using phone numbers and IP addresses as public identities to deploy secure communication between mobile phones and laptops. In the same way, the described scheme can be also used for Wi-Fi pairing.

Let us suppose that two users A and B want to securely share information through the Bluetooth or Wi-Fi of their smartphones and laptops. Assuming the existence of a TTP playing the role of the PKG server, which can register them as users of the system and provide them with corresponding private keys matching their public identities, our protocol is divided into two main phases, whose main characteristics are shown in Figures 1 and 2:

1) Pairing phase. When the pairing phase begins, following the NFC forum recommendation, both devices transparently share their Bluetooth addresses, device names, and any other relevant information needed for the pairing process in an NFC Data Exchange Format (NDEF) message according to the Extended Inquiry Response (EIR) format specified in the Bluetooth Core Specification. Another important piece of information formed by the phone number or IP address is also shared within this phase because it is used as public identity in the scheme. Once finished this initial data transmission, the Bluetooth channel is already set up so that any information can be securely exchanged between the devices.

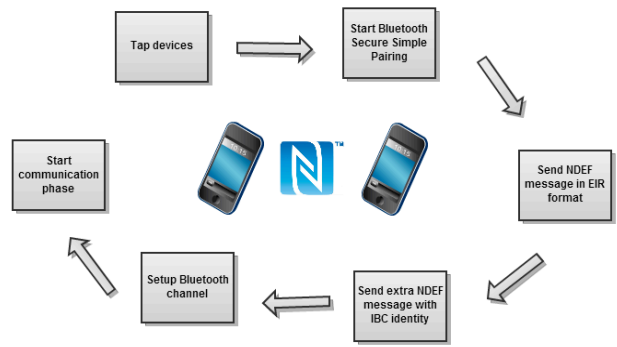

**Figure 1. Pairing Phase.**

2) Communication phase. Once both devices have been paired, this phase begins. As aforementioned, it is important to establish a real secure and trustworthy channel due to the known vulnerabilities of the Bluetooth cipher. IBC is a perfect approach to this problem so it is used in the proposed protocol to guarantee confidentiality, integrity and authenticity. Depending on the size of the data to be shared, the initiating user can decide to send information directly by using an IBC cipher or signcryption algorithm, and/or other cryptographic schemes. For instance, in order to avoid a drastic increase in processing time, a session key is agreed for sharing large amounts of data through an efficient symmetric encryption such as AES, by using an identity-based key agreement protocol like the ones used in [8,24]. On the other hand, to prevent an increase in implementation complexity, an elegant but powerful Diffie-Hellman protocol is run by using the IBC algorithm and by generating random keys through the inspection of data with devices sensors such as gyroscope, accelerometer, compass or GPS location. These methods allow obtaining enough randomness to establish a secure session key to be used in this communication phase.

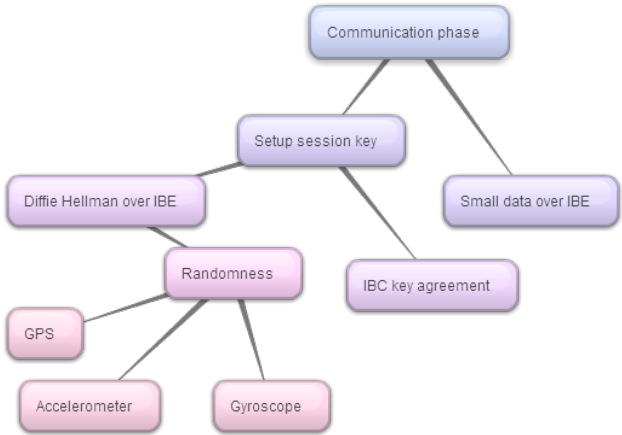

**Figure 2. Communication Phase.**

The IBE algorithm used in the communication phase of our implementation essentially follows the description of Boneh-Franklin scheme [5] based on bilinear pairings, with a few variations. Thus, the used scheme is divided into the following four main stages:

1. Setup stage. The PKG maps string identities to points on an elliptic curve. It sets its own public key $Pu_{PKG}$ as the point $sP$ where $s$ is a random number in $Z^*_q$, and $P$ is an arbitrary point in $E/F_p$ of order $q$. It also chooses a cryptographic hash function $H_1: F_p^2 \rightarrow \{0,1\}^n$ for some $n$, and another cryptographic hash function $H_2: \{0,1\}^* \rightarrow F_p$. The public system parameters are $\langle p,n,P,Pu_{PKG},H_1,H_2 \rangle$ while the private master key is $s \in Z_q$.

2. Extract stage: For any given string $ID \in \{0,1\}^*$, the PKG builds the public key corresponding to such an ID, $Pu_{ID}=H_2(ID)$, a point in $E/F_p$ mapped from ID, and the private key $Pr_{ID}=sPu_{ID}$.

3. Encrypt stage: In order to encrypt and send a message to another user with identity ID, any user chooses a random $r \in Z_q$, and computes the ciphertext $C=\langle rP,M \oplus H_1(g_{ID}^r) \rangle$ where $g_{ID}=e(Pu_{ID},Pu_{PKG}) \in F_p^2$.

4. Decrypt stage: A user with identity ID decrypts the ciphertext $C=\langle U,V \rangle$ encrypted with its public key, by using its private key $Pr_{ID}$ so that it gets the original message $M=V \oplus H_1(e(Pr_{ID},U))$.

As in other IBE schemes, if the PKG is compromised, all the messages protected over the entire lifetime of the public-private key pair used by that server are also compromised.

The implementations of the scheme that we have made are based on a particular family of supersingular elliptic curves $E/F_p$, known as type-1 curves. In the designed IBE, each user with identity ID calculates its public key $Pu_{ID}$ from its phone number or IP address, and the PKG calculates the corresponding private key $Pr_{ID}$ from that public key by using its master key, as explained in the aforementioned extract stage. In the encrypt stage, the sender A chooses a Content Encryption Key (CEK) and uses it to encrypt the message M, encrypts the CEK with the public key $Pu_B$ of the receiver B, and then sends to B both the encrypted message C and the encrypted CEK. Finally, in the decrypt stage, B uses its private key $Pr_B$, which may be securely obtained from the PKG after a successful authentication, in order to decrypt the CEK, which is then used to decrypt the encrypted message C.

Regarding the elliptic curves used in the IBE schemes, the implementations of the application are based on type-1 curves of the form $y^2=x^3+1$ defined over $F_p$ for primes p congruent to 11 modulo 12, mainly because these curves can be easily generated at random. The algorithm uses both the group $E(F_p)$ of points (x,y) satisfying the curve equation with affine coordinates x,y in $F_p$, and the group $E(F_p^2)$ of points (x,y) satisfying the curve equation with affine coordinates x,y in $F_p^2$, and with corresponding Jacobian projective coordinates X,Y,Z also in $F_p^2$.

Under the aforementioned conditions, the Tate pairing e takes as input two points P and Q in $E(F_p)$, and produces as output an integer in $F_p^2$. We chose the Tate pairing as the core of the implemented IBE because it is a map that is hard to invert, but efficiently computable using Miller's algorithm and Solinas primes, thanks to its linearity in both arguments. However, when implementing it with smartphones, we found that although pairing in the projective system is faster than in the affine system, the cost is still very high in some cases, so pre-computation of some intermediate results, such as lambda for the sum of points, was incorporated as solution to speed-up the whole process.

According to the used Boneh-Franklin identity-based cryptosystem, during the setup stage the PKG computes the private master key as an integer s, and the public system parameters such as a prime p, two hash functions $H_1$ and $H_2$, a point P in the elliptic curve $E(F_p)$ and another point obtained by multiplying s times the point P. Afterwards, from the set of public parameters, each public key is obtained by each user with identity ID as a point $Pu_{ID}$ in $E(F_p)$. During the extract stage, the PKG returns to each applicant user with phone number ID, its private key in the form of another point in $E(F_p)$. In this way, the encryption of a message M with the public key $Pu_{ID}$ of a receiver whose phone number is ID, involves both a multiple rP of the point P (being r a randomly chosen integer) and the XOR of M and the hash of the r-power of the pairing result on both public keys. In this way, the decryption of C is possible by adding to its second element, the hash of the pairing result of the private key and its first element.

The proposed protocol is simple, secure, transparent and extensible to be used with any other communication technologies such as WiFi-direct, so that the only necessary change is in the NFC pairing details.

During the implementation of the proposed IBE scheme we have taken advantage of many primitives included in the pairing-based cryptographic library [21], which is a free C library (released under the GNU Lesser General Public License) and built on the GMP library, because such a library is fully functional and ready-to-use.

A few details concerning the technical features of the realization of the proposed application are provided below. In particular, a preliminary implementation has been developed for laptops and smartphones in the Windows Phone and iOS platforms.

The used version of the Windows Phone platform has been Windows Phone 8. Thus, we have used the Windows Phone 8 SDK, together with Visual Studio Ultimate 2012. All tests have been run either in the WP emulator or in a Nokia Lumia 920 smartphone. In particular, the specifications for the used smartphone are:

- Processor name: Qualcomm Snapdragon™ S4
- Processor type: Dual-core 1.5 GHz
- RAM: 1 GB
- Bluetooth: Bluetooth 3.0

The external libraries used in the Windows implementation have been:

- 32feet.NET.Phone, a shared source library to simplify bluetooth development.
- Bouncy Castle, a lightweight cryptography API to make up for the shortages of the WP cryptography API.

The times in the Nokia device are shown in Table 2.

**Table 2. Times on Windows Phone**

| Message size (bytes) | Time to encrypt | Time to decrypt |
|---|---|---|
| 128 | 7497,198 ms | 7368,289 ms |
| 512 | 7498,221 ms | 6998,858 ms |

On the other hand, the technical specifications of the iOS platform used in the initial implementation have been:

- Architecture: Intel Core i5 (64 bits)
- CPU Frequency: 2.4 Ghz
- L1 - L2/L3 Cache: 32k/32k x2 - 256k x2, 3MB
- RAM Memory: 8GB 1333MHz DDR3

The implementation in iOS has been carried out in C language, and the compiler used was i686-apple-darwin11-llvm-gcc-4.2 running in a Mac OS X 10.7.5.

The experiments in iOS have consisted of the encryption and decryption using the proposal. The average time corresponding to 10 runs has been 62,7 milliseconds. Besides, we have performed the activity monitor for 100 runs and the obtained results have been:

- Architecture: Intel Core i5 (64 bits)
- Threads: 1
- Real Mem: 793,00KB
- Virtual Mem: 76,52 MB
- %CPU: 100,2
- CPU Time: 0.6263084

Furthermore, we have started to perform several preliminary analyses on an iPhone 3GS whose main characteristics are:

- Architecture: Armv7-A
- CPU Frequency: 600 MHz
- Cache L1I/L1D/L2: 16 Kb/16 Kb/256 Kb
- RAM: 256 MB

The different implementations of the prototype are still being improved but, as shown above, the preliminary results that have been obtained till now are promising.

## 4. ACKNOWLEDGMENTS
Research supported by Spanish MINECO and European FEDER Funds under projects TIN2011-25452 and IPT-2012-0585-370000.

## 5. CONCLUSIONS AND FUTURE WORK
It is widely clear that the use of smartphones and laptops has overtaken desktop computers and is growing everyday in all types of environments, including in the home. Given the different wireless technologies that are available in these devices, new ways or mechanisms to secure communications must be developed keeping them simple, efficient, energy-saving and still functional and practical. This is the key objective of this work, where Identity-Based Cryptography allows implementing secure, useful and efficient schemes in smart mobile devices.

The main goal of this paper has been to propose an IBC-based security protocol whose implementation is feasible on mobile devices. In particular, here a new Bluetooth and Wi-Fi secure protocol, using NFC for pairing and IBE for communications has been proposed.

Current work in progress is aimed at implementing the proposal in different testing environments and platforms, and at doing an in-depth analysis of its security. Our future research will focus on improving the IBE-based scheme and developing extensions such as signcryption and possible applications in everyday life.